\documentstyle[epsfig]{article}
\title{From the solutions of diffusion equation \\
to the solutions of subdiffusive one}
\author{Tadeusz Koszto{\l}owicz}
\date {\it Institute of Physics, \'Swi\c{e}tokrzyska Academy,\\ 
ul. \'Swi\c{e}tokrzyska 15, 25-406 Kielce, Poland,\\ 
e-mail: tkoszt\verb+@+pu.kielce.pl}

\begin{document}
\maketitle
\begin{abstract}
Starting with the Green's functions found for normal diffusion, we construct 
exact time-dependent Green's functions for subdiffusive equation with 
fractional time derivatives. The Green's functions satisfy the boundary 
conditions involving a linear combination of fluxes and concentrations. 
The method is particularly useful to calculate the concentration profiles 
in a multi-part system where different kinds of transport occurs in each 
part of it. As an example, we find the solution of subdiffusive equation 
for the system composed of two parts with normal diffusion and subdiffusion. 
\end{abstract}

\section{Introduction}

Subdiffusion occurs in many physical systems such as amorphous semiconductors, 
porous media \cite{bg,mk} or gel solvents \cite{kd,dwk}. The subdiffusion 
is defined by the relation 
	\begin{equation}
	\label{a}
\left< \Delta x^{2}\right> \sim t^{\alpha}
	\end{equation} 
with $0<\alpha<1$, where $\left< \Delta x^{2}\right>$ denotes the mean 
square displacement of the transported particle, $t$ is the time. The case 
of $\alpha=1$ corresponds to the normal diffusion. To obtain the concentration 
profiles ($C_{\alpha}$) of transported substance in subdiffusive systems one 
uses the subdiffusion equation with fractional time derivative \cite{mk,com}
	\begin{equation}
	\label{se}
EQ_{\alpha}\left( C_{\alpha}\right) \equiv \left(\frac{\partial}{\partial t} -D_{\alpha}\frac{\partial^{1-\alpha}}{\partial t^{1-\alpha}}\frac{\partial^{2}}{\partial x^{2}}
\right) C_{\alpha}\left( x,t\right) =0\;,
	\end{equation}
with the Riemann-Liouville fractional derivative defined as (for $\alpha>0$)
	\begin{displaymath}
\frac{\partial^{\alpha}f(t)}{\partial t^{\alpha}}=\frac{1}{\Gamma (n-\alpha)}\frac{\partial^{n}}{\partial t^{n}}\int^{t}_{0}dt'\frac{f(t')}{(t-t')^{1+\alpha -n}}\;,
	\end{displaymath}
where $n$ is the smallest integer larger than $\alpha$, $D_{\alpha}$ is the 
subdiffusion coefficient measured in the units $(\frac{m^{2}}{s^{\alpha}})$. 
While there is a huge literature about the normal diffusion in a variety of 
systems (see e.g. \cite{c} and references therein), the subdiffusion is 
satisfactorily described in a few cases only. The procedure of solving 
the equation (\ref{se}) appears to be rather complicated. Below we present 
a practical method to calculate the solutions of the equations describing 
in a multi-part system where diffusion and subdiffusion occur. The starting 
point is the solution corresponding to the analogous multi-part system 
where only normal diffusion is present. 

Till now homogeneous systems have been mostly studied. However, the multi-part 
systems discussed here, in particular the membrane systems, are important for 
several fields of technology and biophysics. We mention here as examples 
the membranes used as micro filters and the (sub-)diffusion processes in 
living cells which are crucial for the cell physiology. There are a few 
attempts described in the literature to obtain solutions of eq. (\ref{se}) 
from the ones corresponding to the normal diffusion. A powerful method seems 
to be mapping of the solutions of diffusion equation $C_{1}(x,t)$ into the 
subdiffusive ones $C_{\alpha}(x,t)$ by means of the relation 
\cite{bmk,b,so,kmk}
	\begin{equation}
	\label{map}
C_{\alpha}(x,t)=\int_{0}^{\infty}n(u,t)C_{1}(x,u)du\;,
	\end{equation}
where the Laplace transform of the function $n$ is 
	\begin{equation}
	\label{sb2}
\hat{n}\left( u,s\right) =\frac{D_{1}}{D_{\alpha}s^{1-\alpha}}e^{-\frac{D_{1}s^{\alpha}}{D_{\alpha}}u},
	\end{equation}
(throughout this paper the index $\alpha$ is assigned to the functions corresponding to subdiffusion; the index $1$ refers to the normal diffusion). Another method is to `rescale' the time as $t\rightarrow t^{\alpha}$ \cite{lm}. Using the substitution one obtains the solutions of normal diffusion equation with diffusion coefficient depending on time. However, both procedures have been used to the homogeneous system only and it is unclear how to apply these approaches to a multi-part system, where the various kinds of diffusion occur (e.g. a porous material, where subdiffusion proceeds, is adjoined to a normal diffusive medium). Thus, a more general procedure is needed. 

	In this paper we discuss a method of constructing solutions of the 
one-dimensional subdiffusive equation with given boundary conditions starting 
with those of normal diffusion, which can be obtained rather easily. Our method 
gives the Green's functions (GF) for subdiffusive equation in the form of 
series of the Fox functions which is convenient for numerical calculations. 
The method splits into two step. In the first one the Laplace transform of 
the subdiffusive Green's function is obtained from the normal diffusion 
counterpart, and in the second one the Laplace transform is inverted. In 
contrast to the previous papers \cite{bmk,b,so,kmk,lm}, we include boundary 
conditions into considerations. As an illustration of our method, we find 
the solutions of eq. (\ref{se}) for the system consisting of two parts where 
the normal diffusion and subdiffusion occur, respectively.

We restrict our considerations to the case of boundary conditions which 
are given in the form of linear combination of concentrations and fluxes. 
We note that this is not a serious restriction as the linear boundary 
conditions often occur in normal diffusion problems (and can be adapted 
for the subdiffusion cases). For example, for the system with fully 
reflecting, fully absorbing \cite{c}, and partially absorbing wall \cite{rn}, 
respectively, the boundary conditions read: $J(x_{M}^{-},t)=0$, 
$C(x_{M}^{-},t)=0$, and $J(x_{M}^{-},t)=\gamma C(x_{M}^{-},t)$, where 
$x_{M}$ denotes a position of the wall. For the partially permeable wall 
there have been used two different boundary conditions: 
$C(x_{M}^{-},t)=\gamma C(x_{M}^{+},t)$ \cite{kpr,dkwm} or 
$J(x_{M}^{-},t)=\lambda (C(x_{M}^{-},t)-C(x_{M}^{+},t))$ \cite{koszpa}, 
which both  should be supplemented by the continuity of the flux 
$J(x_{M}^{-},t)=J(x_{M}^{+},t)$. 

\section{The method}

Since the Green's function for normal diffusion is mapped onto the subdiffusion within the method presented below, we first discuss the connection between the equation of normal diffusion and the subdiffusive equation. 

\subsection{Normal diffusion}

We start with the normal diffusion equation
	\begin{equation}
	\label{ne}
EQ_{1}\left( C_{1}\right) \equiv \left(\frac{\partial}{\partial t} -D_{1}\frac{\partial^{2}}{\partial x^{2}} \right) C_{1}( x,t) =0\;,
	\end{equation}
and assume that each of the boundary condition involves a linear combination of fluxes and concentrations. When the system under consideration is divided into two parts by the membrane located at $x=x_{M}$, the boundary condition (BC) at $x_{M}$ is assumed to be of the form
	\begin{eqnarray}
	\label{bcn}
L_{1}(C_{1}) & \equiv & \delta_{1}C_{1}(x_{M}^{-},t)+
\delta_{2}C_{1}(x_{M}^{+},t)+
\delta_{3}J_{1}(x_{M}^{-},t) \nonumber \\
 & & +\; \delta_{4}J_{1}(x_{M}^{+},t)=0\;,
	\end{eqnarray}
where the flux $J_{1}$ is given as
	\begin{equation}
	\label{fn}
J_{1}(x,t)=-D_{1}\frac{\partial C_{1}(x,t)}{\partial x}\;,
	\end{equation}
and $\delta_{i}$ ($i=1,2,3,4$) are the parameters independent of $t$. 
The solution of the equation (\ref{ne}) can be found due to the following integral formula
	\begin{displaymath}
C_{1}(x,t)=\int G_{1}(x,t;x_{0})C_{1}(x_{0},0)dx_{0}\;,
	\end{displaymath}
where $C_{1}(x_{0},0)$ is the initial concentration, and the Green's function 
$G_1$ is defined as a solution of the equation (\ref{ne})
with the initial concentration
	\begin{equation}
	\label{ic}
G_{1}(x,0;x_{0})=\delta(x-x_{0})\;,
	\end{equation}
with the appropriate boundary conditions. Since the diffusion equation is 
of the second order with respect to the space variable one needs two 
boundary conditions at the discontinuity of the system. When the system 
is infinite, the additional BC are needed at $x\rightarrow \pm \infty$ 
and they are usually of the form 
$G_{\alpha}(x,t;x_{0})\mid_{x\rightarrow\infty}=0$ and 
$G_{\alpha}(x,t;x_{0})\mid_{x\rightarrow -\infty}=0$, which can be treated 
as special cases of eq. (\ref{bcn}) ($\delta_{1}=1$, 
$\delta_{2}=\delta_{3}=\delta_{4}=0$ for $x\rightarrow -\infty$, and 
$\delta_{2}=1$, $\delta_{1}=\delta_{3}=\delta_{4}=0$ for 
$x\rightarrow\infty$, respectively).

One usually solves the normal diffusion equation by means of the Laplace 
transform (LT) \cite{c,cj}
	\begin{displaymath}
\hat{C}(x,s)=\int_{0}^{\infty}e^{-st}C(x,t)dt\;.
	\end{displaymath}
LT of the equation (\ref{ne}), boundary conditions (\ref{bcn}), and flux 
(\ref{fn}) are, respectively,
	\begin{equation}
	\label{ltne}
EQ_{1}(\hat{C}_{1})\equiv s\hat{C}_{1}(x,s)-D_{1}\frac{d^{2}\hat{C}_{1}(x,s)}{d^{2}x}-C_{1}(x,0)=0 \;,
	\end{equation}
	\begin{equation}
	\label{ltbc}
L_{1}(\hat{C}_{1}) \equiv \delta_{1}\hat{C}_{1}(x_{M}^{-},s)+
\delta_{2}\hat{C}_{1}(x_{M}^{+},s)+
\delta_{3}\hat{J}_{1}(x_{M}^{-},s)+\delta_{4}\hat{J}_{1}(x_{M}^{+},s)=0 \;,
	\end{equation}
and
	\begin{equation}
	\label{ltfn}
\hat{J}(x,s)=-D_{1} \frac{d\hat{C}_{1}(x,s)}{dx} \;.
	\end{equation}
The general solution of eq. (\ref{ltne}) with the initial condition (\ref{ic}) is \cite{cj}
	\begin{eqnarray}
	\label{ltnfg}
\hat{G}_{1}\left( x,s;x_{0}\right) & = & A_{1}(s) e^{x\sqrt{s/ D_{1}}} + B_{1}(s) e^{-x\sqrt{s/D_{1}}} \nonumber \\ & & +\frac{1}{2\sqrt{D_{1}s}} e^{-\mid x-x_{0}\mid \sqrt{s/ D_{1}}}\;.
	\end{eqnarray}
where $A_{1}(s)$ and $B_{1}(s)$ are determined by the boundary conditions. 

\subsection{Subdiffusion}

The subdiffusive flux generated by the gradient of the concentration $C_{\alpha}$ is given by the formula 
	\begin{displaymath}
J_{\alpha}(x,t) =-D_{\alpha}\frac{\partial^{1-\alpha}}{\partial t^{1-\alpha}} \frac{\partial C_{\alpha}(x,t)}{\partial x}\;,
	\end{displaymath}
which combined with the continuity condition
	\begin{displaymath}
\frac{\partial C_{\alpha}\left( x,t\right)}{\partial t}+\frac{\partial J_{\alpha}\left(x,t \right)}{\partial x}=0\;,
	\end{displaymath}
provides the subdiffusion equation (\ref{se}).
The boundary condition at $x=x_{M}$ is assumed to be of the form fully 
analogical to that of the normal diffusion (\ref{bcn}) i.e.
	\begin{eqnarray}
	\label{sbc}
L_{\alpha}(C_{\alpha}) & \equiv & \delta_{1}C_{\alpha}(x_{M}^{-},t)+
\delta_{2}C_{\alpha}(x_{M}^{+},t)+
\delta_{3}J_{\alpha}(x_{M}^{-},t) \nonumber \\ & & +\; \delta_{4}J_{\alpha}(x_{M}^{+},t)=0.
	\end{eqnarray}
The Laplace transform of the subdiffusion equation, subdiffusive flux \cite{z} and boundary conditions are, respectively,
	\begin{equation}
	\label{ltse}
{EQ}_{\alpha}(\hat{C}_{\alpha}) \equiv s\hat{C}_{\alpha}(x,s)-s^{1-\alpha}D_{\alpha}
\frac{d^{2}\hat{C}_{\alpha}(x,s)}{dx^{2}}-C_{\alpha}(x,0)=0\;,
	\end{equation}
	\begin{equation}
	\label{ltsf}
\hat{J}_{\alpha}(x,s)=-D_{\alpha}s^{1-\alpha}\frac{d\hat{C}_{\alpha}(x,s)}{dx}\;,
	\end{equation}
	\begin{eqnarray}
	\label{ltsbc}
L_{\alpha}(\hat{C}_{\alpha}) & \equiv & \delta_{1}\hat{C}_{\alpha}(x_{M}^{-},s)+
\delta_{2}\hat{C}_{\alpha}(x_{M}^{+},s)+
\delta_{3}\hat{J}_{\alpha}(x_{M}^{-},s) \nonumber \\
  &  & +\; \delta_{4}\hat{J}_{\alpha}(x_{M}^{+},s)=0\;.
	\end{eqnarray}
The solution of eq. (\ref{ltse}) for the initial condition $C_{\alpha}(x,0)=\delta(x-x_{0})$,
is computed as
	\begin{eqnarray}
	\label{ltsfg}
\hat{G}_{\alpha}\left( x,s;x_{0}\right) & = & A_{\alpha}(s) e^{x\sqrt{s^{\alpha}/D_{\alpha}}} + B_{\alpha}(s) e^{-x\sqrt{s^{\alpha}/D_{\alpha}}} \nonumber \\ & + & \frac{1}{2\sqrt{D_{\alpha}}s^{1-\frac{\alpha}{2}}}
e^{-\mid x-x_{0}\mid \sqrt{s^{\alpha}/D_{\alpha}}},
	\end{eqnarray}
where the coefficients $A_{\alpha}(s)$ and $B_{\alpha}(s)$ are determined, as previously, by the boundary conditions (\ref{ltsbc}). 

\subsection{From the diffusion to subdiffusion equation}

Let us note that substitution
	\begin{equation}
	\label{sb}
D_{1}=D_{\alpha}s^{1-\alpha}
	\end{equation}
maps the Laplace transforms of normal diffusion equation $EQ_{1}\big(\hat{G}_{1}\big)=0$ (\ref{ltne}), and flux $\hat{J}_{1}\big(x,s;x_{0}\big)$ (\ref{ltfn}) onto the subdiffusion ones $EQ_{\alpha}\big(\hat{G}_{\alpha}\big)=0$ (\ref{ltse}) and $\hat{J}_{\alpha}\big(x,s;x_{0}\big)$ (\ref{ltsf}), respectively. Assuming that the coefficients $\delta_{1}$, $\delta_{2}$, $\delta_{3}$, and $\delta_{4}$ are the same for the normal diffusion and for subdiffusion boundary conditions (\ref{ltbc}) and (\ref{ltsbc}), the substitution (\ref{sb}) transforms also the diffusion boundary condition $L_{1}\big(\hat{G}_{1}\big)=0$ (\ref{ltbc}) into the subdiffusion one $L_{\alpha}\big(\hat{G}_{\alpha}\big)=0$ (\ref{ltsbc}). Since the Green's functions are uniquely determined by the subdiffusion equation, boundary conditions and initial condition (which are assumed to be the same for normal diffusion and subdiffusion), it is clear that the substitution (\ref{sb}) transforms the Green's function for normal diffusion (\ref{ltnfg}) into the Green's function for subdiffusion (\ref{ltsfg}). Comparing eq. (\ref{ltnfg}) and eq. (\ref{ltsfg}) we deduce that the coefficients $A_{\alpha}(s)$ and $B_{\alpha}(s)$ can be obtained from $A_{1}(s)$ and $B_{1}(s)$ by means of the replacement (\ref{sb}).

\subsection{Inverse Laplace transform} 

As the Green's function is given by eq. (\ref{ltsfg}), with the functions $A_{\alpha}(s)$ and $B_{\alpha}(s)$ of rather complicated structure (as we will see in Sec. 3.2), the calculation of the inverse Laplace transform is far not simple. Therefore, we expand the function into the following series 
	\begin{eqnarray*}
\hat{G}_{\alpha}\big(x,s;x_{0}\big) & = & \sum_{n=0}^{\infty}a_{n}s^{\beta+n\nu}e^{x\sqrt{s^{\alpha}/ D_{\alpha}}}
+\sum_{n=0}^{\infty}b_{n}s^{\gamma+n\sigma}e^{-x\sqrt{s^{\alpha}/ D_{\alpha}}}
\nonumber \\ & + & \frac{1}{2\sqrt{D_{\alpha}}s^{1-\frac{\alpha}{2}}}e^{-\mid x-x_{0}\mid \sqrt{s^{\alpha}/ D_{\alpha}}} \;.
	\end{eqnarray*}
When the Laplace transform is in such a form, the inverse transformation can be easily calculated, using the formula derived by means of the method based on the Mellin transform \cite{s} (for details of the derivation see the Appendix)
	\begin{equation}
	\label{ge}
f_{\nu ,\beta}(t;a)\equiv L^{-1}\big(s^{\nu}e^{-as^{\beta}}\big)=\frac{1}{\beta 
a^{\frac{1+\nu}{\beta}}}
H_{1\;1}^{1\;0}\left(\frac{a^{\frac{1}{\beta}}}{t}\mid^{\;\; 1\;\;\;\;1}_{\frac{1+\nu}{\beta}\;
\frac{1}{\beta}}\right)\;,
	\end{equation}
where $H$ denotes the Fox function, $a,\beta>0$, and the parameter $\nu$ is not limited. 
	Due to the formula (\ref{ge}), we obtain the Green's function as series of Fox functions. As shown in the Appendix, the function $f_{\nu, \beta}$ is given by the formula 
	\begin{equation}
	\label{p}
f_{\nu, \beta}\left( t;a\right) =-\frac{1}{\pi t^{1+\nu}}\sum_{k=0}^{\infty}\frac{\sin \left [\pi\left (k\beta+\nu \right )\right]\Gamma \left (1+k\beta +\nu\right )}{k!}\left (-\frac{a}{t^{\beta}}\right )^{k}\;.
	\end{equation}
The domain of applicability of eq. (\ref{p}) appears to be limited \cite{s,sgg}.
The Fox function (\ref{ge}) is analytical and has its series expansion 
(\ref{p}) for every non-zero values of the variable 
$\frac{a^{\frac{1}{\beta}}}{t}$ when $\frac{1}{\beta}-1>0$. Otherwise, the 
additional restrictions should be imposed. For example, when $\beta =1$, 
the equation (\ref{p}) is still valid under the condition $a<t$. In the 
following we consider the cases where $\beta<1$, so we will not further 
discuss the domain of applicability of the formula (\ref{p}).

\section{Green's functions for subdiffusive systems}

To illustrate the considerations of the previous section, we find here 
the Green's functions for the homogeneous system, which are well-known, 
and next we discuss a composite system containing two parts where, 
respectively, the normal diffusion and subdiffusion occur. As far as 
we know, such a system has not been considered yet. Leaving the detailed 
analysis of this system to the next paper, we derive here the 
respective Green's functions.

\subsection{The homogeneous system}

Since the subdiffusive homogeneous system has been already described in detail, we only show here that the procedure presented above gives the same results as that of the paper \cite{mk}.
For the homogeneous system of normal diffusion, the functions $A(s)$ and $B(s)$ from eq. (\ref{ltnfg}) vanish. So, the Laplace transform of GF is
	\begin{equation}
	\label{ltfsr}
\hat{G}_{1}\big(x,s;x_{0}\big) =\frac{1}{2\sqrt{D_{1}s}}
e^{-\mid x-x_{0}\mid \sqrt{s/D_{1}}}\;,
	\end{equation}
and the inverse Laplace transform of (\ref{ltfsr}) gives the Gausian function
	\begin{displaymath}
G_{1}(x,t;x_{0})=\frac{1}{2\sqrt{\pi D_{1}t}}e^{-(x-x_{0})^{2}/4Dt}\\.
	\end{displaymath}
The replacement (\ref{sb}) transforms the GF for normal diffusion into the following function
	\begin{displaymath}
\hat{G}_{\alpha}\big(x,s;x_{0}\big)=\frac{1}{2\sqrt{D_{\alpha}}s^{1-\frac{\alpha}{2}}}
e^{-\mid x-x_{0}\mid \sqrt{s^{\alpha}/ D_{\alpha}}}\;,
	\end{displaymath}	
and due to eq. (\ref{ge}) we obtain 
	\begin{eqnarray}
	\label{ge1}
G_{\alpha}\big(x,t;x_{0}\big) & = & \frac{1}{2\sqrt{D_{\alpha}}}f_{\frac{\alpha}{2}-1,\frac{\alpha}{2}}\left(t;\frac{\mid x-x_{0}\mid}{\sqrt{D_{\alpha}}}\right)\nonumber\\ & = & \frac{1}{\alpha\mid x-x_{0}\mid}
H_{1\;1}^{1\;0}\left(\bigg(\frac{(x-x_{0})^{2}}{D_{\alpha}t}\bigg)^{\frac{1}{\alpha}}\mid^{\;\; 1\;\;\;\;1}_{\;\; 1\;\;\;\frac{\alpha}{2}}\right)\;.
	\end{eqnarray}
Let us note that the Fox function has several different representations equivalent to each other. Using the formula \cite{sgg}
	\begin{eqnarray*}
H_{1\;1}^{1\;0}\left(\bigg(\frac{(x-x_{0})^{2}}{D_{\alpha}t}\bigg)^{\frac{1}{\alpha}}\mid^{\;\; 1\;\;\;\;1}_{\;\; 1\;\;\;\frac{\alpha}{2}}\right) & = & \frac{\alpha}{2}H_{1\;1}^{1\;0}\bigg(\frac{x-x_{0}}{\sqrt{D_{\alpha}t^{\alpha}}}\mid^{\;\; 1\;\;\;\;\frac{\alpha}{2}}_{\;\; 1\;\;\;\;1}\bigg)\\[2mm] & = & \frac{\alpha}{2}\frac{\mid x-x_{0}\mid}{\sqrt{D_{\alpha}t^{\alpha}}}H_{1\;1}^{1\;0}\bigg(\frac{\mid x-x_{0}\mid}{\sqrt{D_{\alpha}t^{\alpha}}}\mid^{\;\; 1-\frac{\alpha}{2}\;\;\;\;\frac{\alpha}{2}}_{\;\;\;\; 0\;\;\;\;\;\;\;1}\bigg)\,
	\end{eqnarray*}
we transform the Green's function (\ref{ge1}) to the one derived in \cite{mk}.

\subsection{The two-part system}

As a second example, we consider a system composed of two parts: $L$ ($x<x_{M}$) where the normal diffusion occurs with the diffusion coefficient $D_{1L}$ and $R$ ($x>x_{M}$) where there is the subdiffusion with the subdiffusion coefficient $D_{\alpha R}$. We assume the following boundary conditions at $x_{M}$
	\begin{equation}
	\label{bc1}
J_{1}\left( x_{M}^{-},t\right) =J_{\alpha}\left( x_{M}^{+},t\right)\;,
	\end{equation}
	\begin{equation}
	\label{bc2}
G_{1}\left( x_{M}^{-},t\right) =G_{\alpha}\left( x_{M}^{+},t\right)\;.
	\end{equation}
To use the procedure presented in the previous section, we temporarily assume that in the region $R$ the normal diffusion occurs with the diffusion coefficient $D_{1R}$. Assuming that $x_{0}<x_{M}$, the solutions of the diffusion equation with the boundary conditions (\ref{bc1}) and (\ref{bc2}) can be written in terms of the Laplace transform as follows
	\begin{eqnarray}
	\label{ltnl}
\hat{G}_{11--}\left( x,s;x_{0}\right) & = & \left(\frac{1}{2\sqrt{D_{1L}s}}-\frac{1}{\sqrt{D_{1L}s}\left(\sqrt{D_{1L}/D_{1R}}+1\right)}\right) e^{-\left( 2x_{M}-x-x_{0}\right) \sqrt{s/ D_{1L}}}\nonumber\\[2mm] & + & \frac{1}{2\sqrt{D_{1L}s}}e^{-\mid x-x_{0}\mid \sqrt{s/ D_{1L}}}\;,
	\end{eqnarray}
for $x<x_{M}$, and
	\begin{equation}
	\label{ltnr}
\hat{G}_{11+-}\left( x,s;x_{0}\right)=\frac{1}{\sqrt{D_{1L}s}+\sqrt{D_{1R}s}} e^{-\left( x_{M}-x_{0}\right) \sqrt{s/ D_{1L}}-(x-x_{M})\sqrt{s/ D_{1R}}}\;,
	\end{equation}
for $x>x_{M}$ (the index $11$ of Green's function is assigned to the system where the normal diffusion occurs in both parts, the index $1\alpha$ corresponds to the system with normal diffusion and subdiffusion, the indices $+$, $-$ refer to the signs of $x-x_{M}$ and $x_{0}-x_{M}$, respectively). 

	In the next step we pass from $11$ to $1\alpha$ functions. Let us note that for the homogeneous system the substitution (\ref{sb}) has been done for the whole system. However, for the composite one the substitution should be performed only with respect to the part where subdiffusion occurs. Thus, 
	\begin{equation}
	\label{sb1}
D_{1R}=D_{\alpha R}s^{1-\alpha}\\,
	\end{equation}
whereas the coefficient $D_{1L}$ remains unchanged. Substituting (\ref{sb1}) in the functions (\ref{ltnl}) and (\ref{ltnr}) we obtain
\begin{eqnarray}
	\label{ltsl}
\hat{G}_{1\alpha --}\left( x,s;x_{0}\right) & = & 
\left(\frac{1}{2\sqrt{D_{1L}s}}-\frac{1}{\sqrt{D_{1L}s}
\left(\sqrt{D_{1L}s^{\alpha -1}/D_{\alpha R}}+1\right)}\right) 
e^{-\left( 2x_{M}-x-x_{0}\right) \sqrt{s/ D_{1L}}}
\nonumber\\[2mm] 
& + & \frac{1}{2\sqrt{D_{1L}s}}e^{-\mid x-x_{0}\mid \sqrt{s/ D_{1L}}}\;,
	\end{eqnarray}
and
	\begin{equation}
	\label{ltsr}
\hat{G}_{1\alpha +-}\left( x,s;x_{0}\right)=\frac{1}{\sqrt{D_{1L}s}+\sqrt{D_{\alpha R}s^{2-\alpha}}} e^{-\left( x_{M}-x_{0}\right) \sqrt{s/ D_{1L}}-(x-x_{M})\sqrt{s^{\alpha}/D_{\alpha R}}}\;.
	\end{equation}
It is easy to see that the functions (\ref{ltsl}) and (\ref{ltsr}) fulfil the Laplace transforms of the boundary conditions (\ref{bc1}) and (\ref{bc2}).  
Expanding these functions into the series, we obtain 
	\begin{eqnarray}
	\label{lts--}
\hat{G}_{1\alpha --}(x,s;x_{0})=\frac{1}{2\sqrt{sD_{1L}}}\left( e^{-\mid x-x_{0}\mid\sqrt{s/ D_{1L}}}+e^{-(2x_{M}-x-x_{0})\sqrt{s/ D_{1L}}}\right)\nonumber \\[2mm] 
+\frac{1}{\sqrt{D_{1L}}}\sum_{n=0}^{\infty}\left(-\sqrt{\frac{D_{\alpha R}}{D_{1L}}}\right)^{n+1}s^{\frac{(1-\alpha)n-\alpha}{2}}e^{-( 2x_{M}-x-x_{0})\sqrt{s/ D_{1L}}}\;,
	\end{eqnarray}
	\begin{eqnarray}
	\label{lts+-}
\hat{G}_{1\alpha +-}(x,s;x_{0}) & = & \frac{1}{\sqrt{D_{1L}}}\sum_{n=0}^{\infty}\sum_{k=0}^{\infty}\left( -\sqrt{\frac{D_{\alpha R}}{D_{1L}}}\right)^{n}\frac{1}{k!}\left(-\frac{x_{M}-x_{0}}{\sqrt{D_{1L}}}\right)^{k} \nonumber \\[2mm]
& & \times s^{\frac{(1-\alpha)n+k-1}{2}}e^{-(x-x_{M})\sqrt{s^{\alpha}/ D_{\alpha R}}}.
	\end{eqnarray}
Using the formula (\ref{ge}), we obtain from eqs. (\ref{lts--}) and (\ref{lts+-}) the Green's functions for the considered system
	\begin{eqnarray}
	\label{sg--}
G_{1\alpha --}(x,t;x_{0})=\frac{1}{2\sqrt{\pi D_{1L}t}}\left(e^{-(x-x_{0})^{2}/ 4D_{1L}t}+e^{-(2x_{M}-x-x_{0})^{2}/ 4D_{1L}t}\right) \nonumber \\[2mm]
+ \frac{1}{\sqrt{D_{1L}}}\sum_{n=0}^{\infty}\left(-\sqrt{\frac{D_{\alpha R}}{D_{1L}}}\right)^{n+1}f_{\frac{(1-\alpha)n-\alpha}{2},\frac{1}{2}}\left(t;\frac{2x_{M}-x-x_{0}}{\sqrt{D_{1L}}}\right),
	\end{eqnarray}
	\begin{eqnarray}
	\label{sg+-}
G_{1\alpha +-}(x,t;x_{0}) & = & \frac{1}{\sqrt{D_{1L}}}\sum_{n=0}^{\infty}\sum_{k=0}^{\infty}\left(-\sqrt{\frac{D_{\alpha R}}{D_{1L}}}\right)^{n}\frac{1}{k!}\left(-\frac{x_{M}-x_{0}}{\sqrt{D_{1L}}}\right)^{k} \nonumber \\[2mm]
& & \times f_{\frac{(1-\alpha)n+k-1}{2},\frac{\alpha}{2}}\left(t;\frac{x-x_{M}}{\sqrt{D_{1L}}}\right).
	\end{eqnarray}
For $x_{0}>x_{M}$ the calculations are very similar to the ones presented above. In this case the Green's functions are
	\begin{eqnarray}
	\label{sg-+}
G_{1\alpha -+}(x,t;x_{0}) & = & \frac{1}{\sqrt{D_{1L}}}\sum_{n=0}^{\infty}\sum_{k=0}^{\infty}\left(-\sqrt{\frac{D_{\alpha R}}{D_{1L}}}\right)^{n}\frac{1}{k!}\left(\frac{x_{M}-x_{0}}{\sqrt{D_{1L}}}\right)^{k} \nonumber \\[2mm]
& & \times f_{\frac{(1-\alpha)n+\alpha k-1}{2},\frac{1}{2}}\left(t;\frac{x_{M}-x}{\sqrt{D_{1L}}}\right)\;,
	\end{eqnarray}

	\begin{eqnarray}
	\label{sg++}
G_{1\alpha ++}(x,t;x_{0}) & = & \frac{1}{2\sqrt{D_{\alpha R}t}}\left[f_{\frac{\alpha}{2}-1,\frac{\alpha}{2}}\left(t;\frac{\mid x-x_{0}\mid}{\sqrt{D_{\alpha R}}}\right)
+f_{\frac{\alpha}{2}-1,\frac{\alpha}{2}}\left(t;\frac{x+x_{0}-2x_{M}}{\sqrt{D_{\alpha R}}}\right)\right]\nonumber\\[2mm] & - & \frac{1}{\sqrt{D_{\alpha R}}}\sum_{n=0}^{\infty}\left(-\sqrt{\frac{D_{\alpha R}}{D_{1L}}}\right)^{n}f_{\frac{(1-\alpha)n+\alpha-1}{2},\frac{\alpha}{2}}\left(t;\frac{x+x_{0}-2x_{M}}{\sqrt{D_{\alpha}}}\right).
	\end{eqnarray}

To illustrate the above results we present two plots of the Green's functions for several values of $\alpha$. In Fig. 1 there are presented the Green's functions (\ref{sg--}) and (\ref{sg+-}) when the initial point $x_{0}$ is placed at the region $L$ (with $x_{0}=-1$). In Fig. 2 there are plots of the Green's functions (\ref{sg-+}) and (\ref{sg++}) when the initial point is located at the region $R$ (with $x_{0}=1$); here $x_{M}=0$. We use the arbitrary units and the values of the parameters are: $D_{1L}=1$, $D_{1R}=0.1$, $D_{\alpha R}=0.1$, and $t=10$. For the double series occurring in the functions (\ref{sg+-}) and (\ref{sg-+}), and for the function $f_{\nu,\beta}$ given by eq. (\ref{p}), there are taken first $20$ terms in each of the series. For the functions (\ref{sg--}) and (\ref{sg++}) given by the single series there are taken first $40$ terms and for the function $f_{\nu, \beta}$ we have taken first $70$ terms in the numerical calculations. We have checked that increasing these numbers does not change our results noticeably.

Let us briefly discuss the main qualitative differences between the Green's functions for normal diffusion and the ones for $\alpha <1$. For the case of $x_{0}<x_{M}$ the inflection point of the Green's functions is located at $x_{M}$ for $\alpha<1$ whereas for the normal diffusion this point is placed in the region $R$ in a finite distance from the border between $L$ and $R$. For the case of $x_{0}>x_{M}$ the differences between the considered Green's functions are very similar to those between the Green's function for a homogeneous diffusive and subdiffusive systems studied in \cite{mk}.

\section{Final remarks}

	The procedure presented above allows one to solve the subdiffusion equations for the non-homogeneous systems such as a membrane system, layered system etc. The procedure uses the fact that the normal diffusion equation is simpler to solve than the subdiffusive one and the relation (\ref{sb}) transforms the Green's functions, diffusive equation and boundary conditions from the normal to the subdiffusive case. Thus, knowing the Laplace transform of $\hat{G}_{1}$ one immediately obtains the Laplace transform of $\hat{G}_{\alpha}$.

Summarizing our considerations, the procedure to obtain the Green's functions for the subdiffusive system with linear boundary conditions is presented as a do list:
	\begin{enumerate}
\item find the Laplace transforms of the Green's functions for the analogical system of the normal diffusion, 
\item obtain the Laplace transform of the Green's functions for subdiffusive system using the replacement (\ref{sb}); when the system is divided into several parts, the replacement should be applied only to the part where the subdiffusion occurs,
\item expand the Laplace transform of Green's function $\hat{G}_{\alpha}$ into the power series of $s^{\nu}$ multiplied by $e^{-as^{\beta}}$,
\item use the formula (\ref{ge}) to calculate the inverse Laplace transform.
	\end{enumerate}
The numerical calculations performed in Sec. 3.2, which give the Green's functions presented in Figs. 1 and 2 show usefulness of the method. It gives the explicitly analytic solutions of subdiffusive equations which are appropriate for numerical calculations. 

At the end we return to the method of mapping the Green's function for normal 
diffusion to the subdiffusive one by the relation (\ref{map}) which has been 
mentioned in the Introduction. Let us observe that substituting the expression
(\ref{sb2}) in eq. (\ref{map}) we obtain
	\begin{eqnarray}
	\label{sb3}
\hat{G}_{\alpha}\left( x,s\right) =\frac{D_{1}}{D_{\alpha}s^{1-\alpha}}\int_{0}^{\infty} e^{-\frac{D_{1}s^{\alpha}}{D_{\alpha}}u}G_{1}\left( x,u\right) \nonumber \\
=\frac{D_{1}}{D_{\alpha}s^{1-\alpha}}\hat{G}_{1}\left( x,\frac{D_{1}}{D_{\alpha}}s^{\alpha}\right).
	\end{eqnarray}
If we apply the replacement (\ref{sb}) to the r.h.s. of eq. (\ref{sb3}), 
we get the identity. So, the procedure presented in this paper is consistent 
with the procedure discussed in \cite{b} for the homogeneous system.

\section*{Acknowledgements}

The author wishes to express his thanks to Stanis{\l}aw Mr\'owczy\'nski for fruitful discussions and critical comments on the manuscript.

\section*{Appendix}

We derive here the relations (\ref{ge}) and (\ref{p}). Let us consider the Laplace transform
	\begin{equation}
	\label{a1}
\hat{f}_{\nu,\beta}\left( s;a\right)\equiv L\left[ f\left( t\right)\right]\left( s\right) =s^{\nu}e^{-as^{\beta}}\\.
	\end{equation}
The Mellin transform 
	\begin{displaymath}
M\left[ f\left( x\right)\right]\left( p\right)\equiv \int_{0}^{\infty}x^{p-1}f\left( x\right) dx\\,
	\end{displaymath}
applied to the relation (\ref{a1}) gives
	\begin{equation}
	\label{a2}
M\left[\hat{f}_{\nu,\beta}\left( s;a\right)\right]\left( p\right) =\frac{1}{\beta a^{\frac{p+\nu}{\beta}}}\Gamma\left(\frac{p+\nu}{\beta}\right)\\.
	\end{equation}
Using the well-known formula 
	\begin{displaymath}
M\left[ L\left[ f\left( t\right)\right]\left( s\right)\right] \left( 1-p\right) =\Gamma\left( 1-p\right) M\left[ f\left( t\right)\right]\left( p\right)\\,
	\end{displaymath}
we obtain
	\begin{equation}
	\label{a3}
f\left( t\right) =M^{-1}\left[\frac{1}{\Gamma\left( 1-p\right)}M\left[ L\left[ f\left( t\right)\right]\left( s\right)\right]\left( 1-p\right)\right]\\.
	\end{equation}
Putting eq.(\ref{a2}) in (\ref{a3}) and using the inverse Mellin transform we get
	\begin{displaymath}
f\left( t\right) =\frac{1}{2\pi i}\int_{c-i\infty}^{c+i\infty}t^{-p}\frac{1}{\beta a^{\frac{1-p+\nu}{\beta}}}\frac{\Gamma\left(\frac{1-p+\nu}{\beta}\right)}{\Gamma\left( 1-p\right)}dp\\.
	\end{displaymath}
Comparing above equation with the definition of the Fox $H$-function
	\begin{eqnarray*}
H_{PQ}^{mn}\left( z\big|
\begin{array}{ccc}
\left( a_{1},A_{1}\right) & \ldots & \left( a_{P},A_{P}\right) \\
\left( b_{1},B_{1}\right) & \ldots & \left( b_{Q},B_{Q}\right)
\end{array}
\right) =\frac{1}{2\pi i}\int_{C}z^{p}\frac{A\left( p\right) B\left( p\right)}{C\left( p\right) D\left( p\right)}dp ,
	\end{eqnarray*}
where $A\left( p\right) =\prod_{j=1}^{m}\Gamma\left( b_{j}-B_{j}p\right)$, $B\left( p\right) =\prod_{j=1}^{n}\Gamma\left( 1-a_{j}+A_{j}p\right)$, $C\left( p\right) =\prod_{j=1}^{Q}\Gamma\left( 1-b_{j}+B_{j}p\right)$, $D\left( p\right) =\prod_{j=1}^{P}\Gamma\left( a_{j}-A_{j}p\right)$,
and the circle $C$ separates the poles of $A(p)$ and $B(p)$, we obtain
	\begin{displaymath}
H_{PQ}^{mn}=-\sum_{poles \;of A\left( p\right)}{\rm Res}\left(\frac{A\left( p\right) B\left( p\right)}{C\left( p\right) D\left( p\right)}\right)\\.
	\end{displaymath}
To get the poles of $A(p)$ we take the relation $b_{j}-B_{j}p=-k$, with $k=0,1,2,\ldots$, which gives 
	\begin{equation}
	\label{a4}
p_{j,k}=\frac{b_{j}+k}{B_{j}}\\.
	\end{equation}
Using the relation 
	\begin{eqnarray*}
\lim_{p\rightarrow p_{j,k}}\left( b_{j}-B_{j}p+k\right)\Gamma\left( b_{j}-B_{j}p\right)
= \lim_{p\rightarrow p_{j,k}}\frac{\Gamma\left( b_{j}-B_{j}p+k+1\right)}{\left( b_{j}-B_{j}p\right)\left( b_{j}-B_{j}p+1\right)\ldots \left( b_{j}-B_{j}p+k\right)}\\
= \frac{\Gamma\left( 1\right)}{\left( -k\right)\left( -k+1\right)\left( -k+2\right)\ldots \left( -1\right)}=\frac{( -1)^{k}}{k!}\;,
	\end{eqnarray*}
we find 
	\begin{displaymath}
{\rm Res}_{p=p_{j,k}}\left(\frac{A\left( p\right) B\left( p\right)}
{C\left( p\right) D\left( p\right)}\right) =
-\frac{\left( -1\right)^{k}{A'}_{j}\left( p_{j,k}\right) B
\left( p_{j,k}\right)}{B_{j}k!C\left( p_{j,k}\right) 
D\left( p_{j,k}\right)}z^{p_{j,k}}\;,
	\end{displaymath}
where ${A'}_{j}\left( p\right) =
\frac{A\left( p\right)}{\Gamma\left( b_{j}-B_{j}p\right)}$. 
So, we obtain 
	\begin{equation}
	\label{a5}
H_{PQ}^{mn}\left( z\right) =\sum_{j=1}^{m}\frac{1}{B_{j}}
\sum_{k=0}^{\infty}\frac{\left( -1\right)^{k}{A'}_{j}
\left( p_{j,k}\right) B\left( p_{j,k}\right)}{B_{j}k!C
\left( p_{j,k}\right) D\left( p_{j,k}\right)}z^{p_{j,k}}.
	\end{equation}
Additionally, we use the relation 
$\Gamma\left( z\right)\Gamma\left( 1-z\right) =
\frac{\pi}{\sin \left(\pi z\right)}$ which gives 
$$
\frac{1}{\Gamma\left( -\alpha k\right)}=
-\frac{\sin\left(\pi\alpha k\right)\Gamma\left( 1+\alpha k\right)}{\pi}\;.
$$
From the above equation and the formulas (\ref{a4}) and (\ref{a5}) we get 
the relation (\ref{p}).

\pagebreak

\begin{figure}
\centerline{\epsfig{file=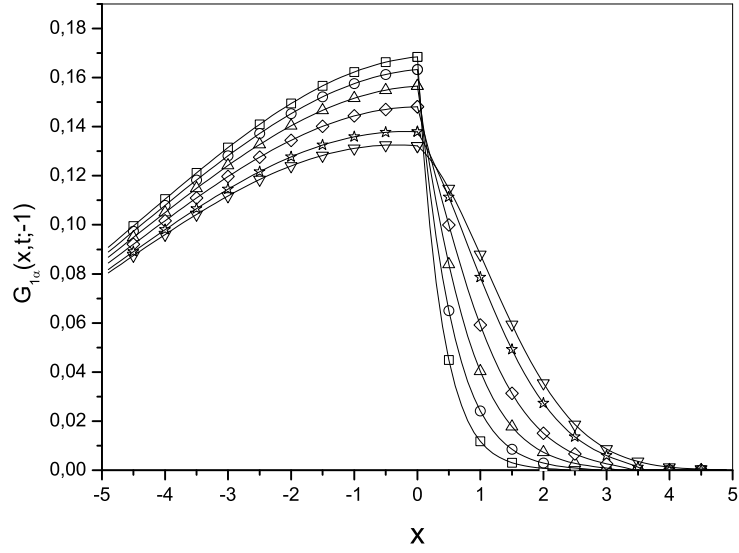,width=110mm}}
\vspace{2mm}
\caption{The Green's functions for the two-part system for $x_{0}=-1$ and several values of the parameter $\alpha$: $\Box$ corresponds to $\alpha=0.1$, $\circ$ to $\alpha=0.3$,  $\triangle$ to $\alpha=0.5$, $\Diamond$ to $\alpha=0.7$, $\star$ to $\alpha=0.9$, and $\bigtriangledown$ to $\alpha=1,0$ (the normal diffusion case).}
\end{figure}

\begin{figure}
\centerline{\epsfig{file=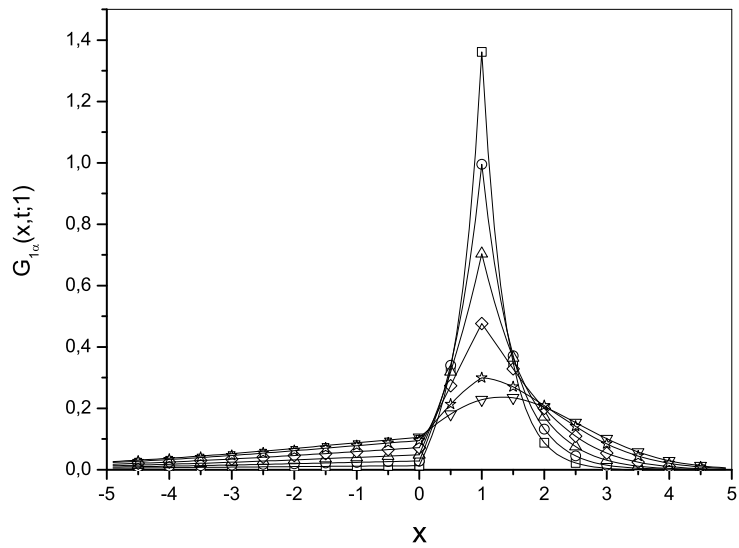,width=110mm}}
\vspace{2mm}
\caption{The Green's functions for the two-part system for $x_{0}=1$ and several values of the parameter $\alpha$ (the description of the symbols is the same as for Fig.1).}
\end{figure}

\end{document}